\documentclass[twocolumn,english,showpacs]{revtex4}

\usepackage{babel,amsmath,amssymb,dcolumn}
\usepackage[dvips]{graphics}
\usepackage[dvips]{graphics}
\newcommand{\bea}{\begin{eqnarray}}
\newcommand{\eea}{\end{eqnarray}}
\newcommand{\bs}{\begin{slide}}
\newcommand{\es}{\end{slide}}
\newcommand{\bi}{\begin{itemize}}
\newcommand{\ei}{\end{itemize}}
\newcommand{\beq}{\begin{equation}}
\newcommand{\eeq}{\end{equation}}
\newcommand{\om}{\omega}

\begin{document}

\title{Quasi-Normal Modes of Brane-Localised Standard Model Fields}
\author{P. Kanti}

\email{panagiota.kanti@durham.ac.uk}

\affiliation{Department of Mathematical Sciences, University of Durham\\
Science Site, South Road, Durham DH1 3LE, United Kingdom}

\author{R. A. Konoplya}

\email{konoplya@fma.if.usp.br}

\affiliation{Instituto de F\'{\i}sica, Universidade de S\~{a}o Paulo \\
C.P. 66318, 05315-970, S\~{a}o Paulo-SP, Brazil}


%


%


%



\begin{abstract}

We present here a detailed study of the quasi-normal spectrum of 
brane-localised Standard Model fields in the vicinity of $D$-dimensional
black-holes. A variety of such backgrounds (Schwarzschild,
Reis\-sner-Nordstr\"om and Schwarzszchild-(Anti) de Sitter) are investigated.
The dependence of the quasi-normal spectra on the dimensionality $D$,
spin of the field $s$, and multipole number $\ell$  is analyzed. 
Analytical formulae are obtained for a number of limiting cases: 
in the limit of large multipole number for Schwarzschild, Schwarzschild-de Sitter and
Reissner-Nordstr\"om black holes,  in the extremal limit of
the Schwarzschild-de Sitter black hole, and in the limit of small
horizon radius in the case of Schwarzschild-Anti de Sitter black holes.
We show that an increase in the number of hidden, extra dimensions results
in the faster damping of all fields living on the brane, and that the localization
of fields on a brane affects the QN spectrum in a number of additional ways,
both direct and indirect.

\end{abstract}

\maketitle


\section{Introduction}

Upon an external perturbation of a black hole background, realized either through the
addition of a field or by perturbing the metric itself, the gravitational system enters
a phase of damping oscillations, or quasi-normal ringing \cite{Nollert1, KS} as it is 
alternatively called. During this phase, the frequency of the field consists of a
real part $\omega_{\rm Re}$,
that drives the field oscillations, and of an imaginary part $\omega_{\rm Im}$, that
cau\-ses the simultaneous damping of these oscil\-lations. The smaller  $\omega_{\rm Im}$ is,
the longer the damping time, therefore certain quasi-normal modes (QNMs)
can dominate the spectrum at very late times after the
initial perturbation, thus governing the dynamical evolution of the black hole. 

The spectrum of quasi-normal modes has been the subject of an intensive study over 
the years, not only for their theoretical interest, but also due to the fact that
their potential experimental detection could lead to the discovery of black holes.
In a 4-dimensional context, they have been both numerically and analytically studied
for a variety of black-hole backgrounds \cite{quasi-4D}. Among the different species
of fields, the quasi-normal modes associated to gravitons, generated by metric
perturbations, were particularly looked at, for the simple reason that gravitational
QNMs originating from astrophysical black holes could potentially be observed with
the help of gravitational wave detectors \cite{Nollert1, KS}. Unfortunately, such
an experimental confirmation has not been obtained up to now. 

A few years ago, the landscape in gravitational physics changed with the formulation
of theories postulating the existence of additional spacelike dimensions in nature
\cite{RS, ADD}. According to these theories, all Standard Model (SM) particles
(scalars, fermions and gauge bosons) are restricted to live on a (3+1)-dimensional
hypersurface -- a brane -- embedded in the higher-dimensional `bulk'. Gravitons
can progagate both on and off the brane, with the same being true for other
particles, like scalars, that carry no charges under the Standard Model gauge group.
This geometrical set-up protects the accurately observed properties of SM fields
from being altered due to the presence of extra dimensions while opening the way
for the study of new gravitational backgrounds.

The theory with Large Extra Dimensions \cite{ADD} predicts the existence of $d$
additional spacelike compact dimensions, all having -- in the simplest case --
the same size $L$. Black holes produced by the gravitational collapse of matter
on the brane would naturally extend off the brane, being gravitational objects.
Whereas macroscopic astrophysical black holes extend mainly along the usual three,
non-compact spatial dimensions, thus being effectively 4-dimensional objects,
microscopic black holes with a horizon radius $r_H \ll L$ would virtually live
in a non-compact spacetime with $D=4+d$ dimensions in total. Higher-dimensional
generalizations of 4-dimensional black hole solutions \cite{Tangherlini, Myers},
derived previously, came back in the spotlight, and the study of the QNMs
associated with these higher-dimensional backgrounds soon became the subject
of a renewed research activity \cite{quasi-D}. 

One of the most exciting predictions of the theory with Large Extra Dimensions
\cite{ADD} is that such microscopic black holes may be created on ground-based
accelerators during the collision of highly energetic particles with center-of-mass
energy $\sqrt{s}> M_*$ \cite{creation}. The energy scale $M_*$ denotes the fundamental
Planck scale of the higher-dimensional gravitational theory, that becomes effective at
distances $r < L$. This scale can be much lower than the 4-dimensional Planck
scale $M_{Pl}$, even as low as 1 TeV, therefore trans-planckian collisions can
be easily realised at next-generation particle colliders. What is of the 
outmost importance is the fact that these tiny black holes, contrary to what
happens in the case of macroscopic black holes, will be created in our 
neighbourhood in a controlled experiment inside a laboratory; therefore, their
detection, either through the emission of Hawking radiation or the detection of
their QNMs spectrum, will be substantially more favoured. In the latter
case, the spectrum of QN modes associated with SM fields living in our brane
will be the most important of all due to the well-developed techniques for the
detection of fermions and gauge bosons, compared to the ones for the up-to-now
elusive gravitons.

In this work, we attempt to fill the gap in the existing literature by 
presenting a comprehensive study of the QN modes associated with the brane-localized
SM fields. We will examine a variety of higher-dimensional black hole backgrounds,
namely the Schwarzschild, Reis\-sner-Nordstr\"om and Schwarzschild-(Anti) de Sitter,
all described by the same static, spherically-symmetric line-element with a single
metric function. The spectrum of the QN modes for scalars, fermions and gauge
bosons will be derived in each case, as a function of the dimensionality of
spacetime, the multipole number and additional fundamental parameters
such as the bulk cosmological constant and charge of the black hole. In what
follows, we will be assuming that the black hole is characterized by a mass
$M_{BH}$ that is at least a few orders of magnitude larger than the fundamental
scale of gravity $M_*$, so that quantum corrections can be safely ignored. 
Also, in order to avoid a hierarchy problem, the brane self-energy can be
naturally assumed to be of the order of the fundamental Planck scale $M_*$
and thus much smaller than $M_{BH}$, therefore its effect on the gravitational
background can also be ignored.

In section II, we present the theoretical framework for our analysis and
the equations of motion for the SM fields propagating in the brane background.
Section III focuses on the associated QNM spectrum in the case of an
induced-on-the-brane $D$-dimensional Schwarzschild black hole, and investigates
the dependence of the spectrum on the dimensionality of spacetime, spin of the
particle and multipole number. In Section IV we proceed to consider the QN modes
of SM particles propagating on a brane embedded in a charged $D$-dimensional
Reis\-sner-Nordstr\"om black hole, and the effect of the black-hole charge on
the QN spectrum is examined in detail. The spectra of QNMs for brane-localised
SM fields in the background of a $D$-dimensional Schwarzschild-de Sitter and
Schwarzschild-Anti de Sitter black hole are derived in sections V and VI,
respectively, and the role of the bulk cosmological constant is investigated.
We present our conclusions in section VII.


\section{Master equation for propagation of fields on the brane}

As mentioned above, in this work we will concentrate on spherically-symmetric
higher-dimensional black-hole backgrounds arising under the assumption of the
existence of $d=D-4$ additional, compact spacelike dimensions in nature. A large
variety of such backgrounds may be described by a unique line-element of
the form
\begin{equation}
ds^2 = - h(r)\,dt^2 +\frac{dr^2}{h(r)} + r^2\,d\Omega_{d+2}^2\,,
\label{metric-bulk}
\end{equation}
where $d\Omega_{d+2}^2$ is the area of the  ($d+2$)-dimensional unit sphere
given by
\bea
&~& \hspace*{-0.6cm} d\Omega_{d+2}^2=d\theta^2_{d+1} + \sin^2\theta_{d+1} \,\biggl(d\theta_d^2 +
\sin^2\theta_d\,\Bigl(\,... \nonumber \\
&~& \hspace*{1.9cm} + \sin^2\theta_2\,(d\theta_1^2 + \sin^2 \theta_1
\,d\varphi^2)\,...\,\Bigr)\biggr)\,,
\label{unit}
\eea
with $0 <\varphi < 2 \pi$ and $0< \theta_i < \pi$, for $i=1, ..., d+1$.
The radial-dependent metric function $h(r)$ may now be assumed to take the form
\beq
h(r) = 1-\frac{\mu}{r^{D-3}} \,,
\label{h-fun-Sch}
\eeq
describing a higher-dimensional, neutral black hole for\-med in a flat,
empty space \cite{Tangherlini}, with the parameter $\mu$ rela\-ted to the
ADM mass of the black hole through the relation 
\beq
\mu=\frac{\kappa^2_D M_{BH}}{(D-2)\,\pi^{(D-1)/2}}\,
\Gamma\left[\frac{D-1}{2}\right]\,. \label{mu}
\eeq
Alternatively, it may be taken to be 
\beq
h(r) = 1-\frac{\mu}{r^{D-3}} + \frac{Q^2}{r^{2(D-3)}}\,,
\label{h-fun-RN}
\eeq
in the case of a higher-dimensional, charged black hole \cite{Myers} with $Q^2$ the
electromagnetic charge. Finally, it may be written as
\beq
h(r) = 1-\frac{\mu}{r^{D-3}} - \frac{2 \kappa_D^2\,\Lambda\,r^2}{(D-1) (D-2)}
\label{h-fun-dS}
\eeq
to describe a Schwarzschild-(Anti) de Sitter black hole formed in a ($d+4$)-dimensional
spacetime with the presen\-ce of a (negative) positive cosmological constant $\Lambda$
\cite{Tangherlini}. In the above expressions, $\kappa_D^2=8\pi G=8\pi/M_*^{D-2}$ stands
for the higher-dimensional Newton's constant.

In what follows, we will study all three types of black-hole backgrounds and derive
exact numerical results for the associated quasi-normal modes of all species of Standard
Model (SM) fields: scalars, fermions and gauge bosons. According to the assumptions of
the model \cite{ADD}, all SM fields are localised to a brane embedded in the bulk
(\ref{metric-bulk}). Since a potential observer also lives on the brane, the study of
the QNMs of brane-localised fields is the most phenomenologically interesting one.
To this end, we need first to determine the line-element of the background in which
the brane-localized
modes propagate. This can be found by fixing the values of the additional angular
coordinates, describing the compact $d$ dimensions, to $\theta_i=\pi/2$
\cite{kmr1, kmr2}. This results in the projection of the higher-dimensional
background (\ref{metric-bulk}) onto the brane, and in the induced-on-the-brane
line-element
\begin{equation}
ds^2=- h(r)\,dt^2 + \frac{dr^2}{h(r)} + r^2\,(d\theta^2 + \sin^2\theta\,d\varphi^2)\,.
\label{metric-br}
\end{equation}
As the projection clearly affects only the angular part of the higher-dimensional
line element given in Eq. (\ref{unit}), the metric function $h(r)$ remains unaffected
and is still given by Eq. (\ref{h-fun-Sch}), (\ref{h-fun-RN}) or (\ref{h-fun-dS}).
Note that, although the additional, spacelike dimensions have been projected out, 
the induced-on-the-brane line-element carries an explicit dependence on the
number and content of extra dimensions through the metric function $h(r)$.

The next step in our analysis is the derivation of the equations of motion of the
various species of fields propagating on the brane. This task was performed in
\cite{Kanti} (see also \cite{IOP1}), for fields with spin $s=0, 1/2$ and 1, by
using the Newman-Penrose method \cite{NP, Chandrasekhar}. For a field with spin
$s$, the following standard factorization was employed
\beq 
\Psi_s(t,r,\theta,\varphi)=  e^{-i\om t}\,e^{i m \varphi}\,\Delta^{-s}\,
P_s(r)\,S_{s}(\theta)\,,
\eeq
where $\Delta= h r^2$, and $S_s(\theta)$ are the spin-weighted spherical harmonics
\cite{Goldberg}. As it was shown in \cite{Kanti}, the radial and angular parts of
each equation are decoupled, and take the `master' form 
\bea
&~& \hspace*{-0.8cm} \Delta^{s}\,\frac{d \,}{dr}\,\biggl(\Delta^{1-s}\,
\frac{d P_s}{dr}\,\biggr) + \nonumber \\[1mm] &~& \hspace*{0.5cm}
\biggl(\frac{\om^2 r^2}{h} + 2i s\,\om\,r -\frac{i s \om\,r^2 h'}{h}
- \tilde \lambda \biggr)\,P_s=0\,,
\label{radial1}
\eea
and
\bea
&~& \hspace*{-0.8cm} \frac{1}{\sin\theta}\,\frac{d \,}{d \theta}\,\biggl(\sin\theta\,
\frac{d S_s}{d \theta}\,\biggr)  + \biggl[-\frac{2 m s \cot\theta}
{\sin\theta} \nonumber \\[1mm] &~& \hspace*{1cm}
- \frac{m^2}{\sin^2\theta} + s - s^2 \cot^2\theta + \lambda \biggr]\,S_s=0\,,
\label{angular}
\eea
respectively. Following \cite{Teukolsky}, these Teukolsky-like equations hold for the
upper component $s=|s|$ of all fields with spin $s=0,1/2$ and 1. The constant
$\tilde \lambda$ appearing in the radial equation is related to the angular
eigenvalue $\lambda$ as follows
\beq
\tilde \lambda  \equiv \lambda + 2 s =  \ell (\ell+1) - s(s-1)\,.\label{lambda}
\eeq
By choosing the desired expression for the metric function $h(r)$ from the set of
Eqs. (\ref{h-fun-Sch}), (\ref{h-fun-RN}) and (\ref{h-fun-dS}), Eqs. (\ref{radial1})
and (\ref{angular}), then, describe the motion of a particle with spin $s$
localized on a brane embedded in a higher-dimensional Schwarz\-schild,
Reissner-Nordstr\"om and Schwarz\-schild-(Anti) de Sitter background,
respectively.


\subsection{Alternative form: One-Dimensional wave equation}

For the purpose of calculating the QNMs of brane-localized fields living in a 
general, spherically-symmetric background of the form (\ref{metric-br}),
studying the radial equation (\ref{radial1}) will be adequate. In what
follows, we will rewrite this equation in a more convenient form, namely
in the form of a one-dimensional Schr\"odinger (or wave-like) equation,
and derive the form of the corresponding effective potential. To this end,
we follow the analysis performed in \cite{Khanal}, and define a new radial
function and a new (``tortoise'') coordinate according to
\beq
P_s=r^{2(s-1/2)}\,Y_s\,, \quad \qquad 
\frac{dr_*}{dr}=\frac{1}{h}\,.
\eeq
Then, Eq. (\ref{radial1}) takes the form
\beq
\left(\frac{d^2}{dr_*^2} + \omega^2\right) Y_s + {\cal P}
\left(\frac{d}{dr_*} + i\omega\right) Y_s - {\cal Q}\,Y_s=0\,,
\label{new}
\eeq
where we have defined
\beq
{\cal P}=s \left(\frac{4 h}{r} - \frac{\Delta'}{r^2}\right)\,,
\label{P}
\eeq
and
\beq
{\cal Q}=\frac{h}{r^2}\,\left\{\tilde\lambda - (2s-1)(s-1)\left(
2h-\frac{\Delta'}{r}\right)\right\}\,.
\label{Q}
\eeq
By defining further
\beq
Y_s=h Z_s + 2i\om \left(\frac{d}{d r_*} - i \om\right)Z_s\,,
\eeq
Eq. (\ref{new}) can now be written as a one-dimensional Schr\"o\-dinger equation
\beq
\left(\frac{d^2}{dr_*^2} + \omega^2\right) Z_s = V_s\,Z_s\,.
\label{wave}
\eeq
The effective potential $V_s$ is found to be spin-dependent and given by the expression
\beq
V_{s=1}=\frac{h}{r^2}\,\ell(\ell+1)\,,
\label{spin-1}
\eeq
for spin-1 particles, where we have used Eq. (\ref{lambda}), and
\beq
V_{s=1/2}=h\,k\,\left[\frac{k}{r^2} \mp 
\frac{d}{dr}\left(\frac{\sqrt{h}}{r}\right)\right]\,,
\label{spin-1/2}
\eeq
for spin-1/2 particles, where 
\beq
k=\sqrt{\ell(\ell+1)+1/4}=1,2,3,...\,.
\eeq
For spin-0 particles, the quantity ${\cal P}$ defined in Eq. (\ref{P}) vanishes
trivially, and the equation for $Y_s$ (\ref{new}) automatically takes a wave-like form,
with potential
\begin{equation}
V_{s=0}=h \left[\,\frac{\ell(\ell+1)}{r^2} + \frac{h'}{r}\,\right]\,.
\label{spin-0}
\end{equation}
The above results hold for any $D$-dimensional,
sphe\-ri\-cally-symmetric black hole (Schwarzschild, Reis\-sner-Nordstr\"om
and Schwarzschild-dS/AdS) projected onto the brane, upon choosing the
correct value of the metric function $h(r)$.


\section{Neutral non-rotating black hole: WKB values of Brane QNMs}

For a $D$-dimensional Schwarzschild black hole background, the effective
potentials $V_s$, seen by scalars, fermion and gauge bosons propagating on the
embedded brane, have the form of a positive-definite potential barrier  
that attains its maximum value close to the black hole while it vanishes
at the event horizon ($r_{*} = - \infty$) and spatial infinity
($r_{*} = + \infty$). According to Eq. (\ref{wave}) then, the solution for
all types of fields in these two asymptotic regimes can be written in terms
of incoming and outgoing plane waves. If we write the quasi-normal frequency
as $\omega=\omega_{\rm Re} - i\omega_{\rm Im}$, under the choice of the
positive sign for $\omega_{\rm Re}$, the QNMs for all Standard Model
particles propagating on the brane satisfy the boundary conditions \cite{Nollert1, KS}
\begin{equation}
\Psi_s (r_*) \approx C_\pm \exp( \pm i \omega r_*) \quad
\textrm{as} \quad r_* \rightarrow \pm \infty\,,
\end{equation}
corresponding to purely in-going waves at the event horizon and
purely out-going waves at spatial infinity.

The fact that the effective potentials $V_s$ have the form of a positive-definite
potential barrier also allows us to use the well-known WKB method in order to
find the various QNMs with a significantly good accuracy. The WKB formula for QN
frequencies has the form:
\begin{equation}
i \frac{\omega^2 - V_0}{\sqrt{-2V_0^{\prime\prime}}} - L_2 - L_3 - L_4 -
L_5 - L_6 = n + \frac{1}{2}\,, \label{WKB}
\end{equation}
where $V_0$ is the height of the potential at its maximum point, located at $r_0$,
and $V_0^{\prime\prime}$ its second derivative with respect to the tortoise coordinate
at the same point. The lower terms $L_2$ and $L_3$ can be found in \cite{will}, while the
higher ones $L_4$, $L_5$ and $L_6$ in \cite{konoplyaWKB}. Finally, the symbol $n$
in the above formula denotes the various overtones.

\begin{figure}[ht]
\resizebox{1\linewidth}{!}{\includegraphics*{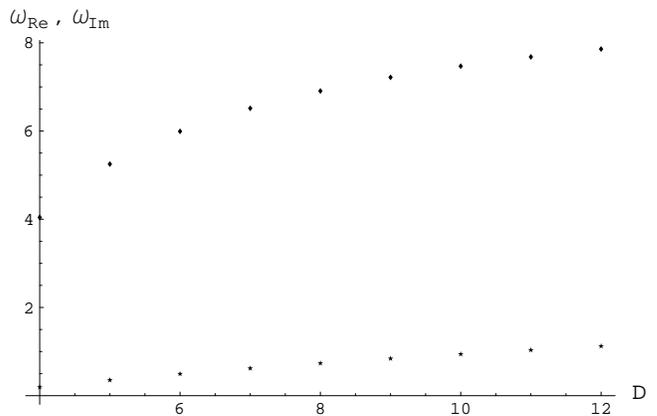}}
\caption{Dependence of real (diamond) and imaginary (star) parts of quasi-normal
modes on spacetime dimensionality $D$ in the eikonal approximation; for each $D$, 
we have chosen $\ell =10$ and set $r_{H}=1$.}   
\label{fig_1}
\end{figure}


By using the above formula, we have calculated the QNMs for all species of SM fields
for various values of the dimensionality $D$ of the spacetime, multipole moment
$\ell$ and overtone number $n$. In Tables I, II and III we display, as
illustrative examples, the QNMs for sca\-lars, fermions and gauge bosons, respectively,
for $D=5$ and $D=6$, and for $n \leq \ell$  up to $\ell=3$. The only numerical
results available in the literature for QN modes of brane-localised fields
are the ones presented in \cite{Berti}, where the case $D=5$ was studied. 
Comparison with their numerical data shows that there is a very good agreement
between our results obtained by using the WKB method and the exact numerical ones,
and that the higher the WKB order is, the better the accuracy. It is known that
the WKB method is accurate for modes with $\ell > n$, therefore 
within the lower overtones, the worst (but still reasonable) precision will be for  
the fundamental scalar mode ($\ell = n = 0$): we have found $0.265391 - 0.497380 i$ 
in 3rd WKB order, and $ 0.256429 - 0.391293 i$ in 6th WKB order, while 
the numerical value is  $0.27339 - 0.41091$, i.e. the relative error
is about 4\% for the real part and 2\% for the imaginary part. For the
fermionic ($k=1, n = 0$) and electromagnetic ($\ell=1, n = 0$) fundamental
mode, the WKB method is expected to be rather accurate since $k, \ell > n$;
indeed, for the electromagnetic mode, for example, we have found 
$0.545243 - 0.331346 i$ in 3rd WKB order, and $0.576508 - 0.313923 i$ in
6th WKB order, while the numerical value is $0.57667 - 0.31749 i$; therefore, the
relative error is less then $0.01$\% for the real part, and about $0.2$\%
for the imaginary part. For higher multipoles, a much better WKB accuracy is
expected. As the reader will note, the two 6-th order WKB values for the modes
$k = n = 1$ and $\ell = n = 1$ for $D = 6$ are absent from Tables II and III:
this is due to the fact that, for particular modes, the WKB method begins to
show bad convergence as $D$ starts taking large values.  

One can find an analytical expression for QNMs in the limit of large multipole
number $\ell$, by making use of the formula (\ref{WKB}) within first order and
expanding in terms of $1/\ell$. In the eikonal approximation -- as this
approximation is called, the following formula for scalar and electromagnetic
perturbations is valid:
\begin{equation} 
\omega = \sqrt{\frac{r_{0}^{D-3} - r_{H}^{D-3}}{r_{0}^{D-1}}}\,\left[\ell
 + \frac{1}{2} - i \sqrt{D-3} \left(n + \frac{1}{2} \right) \right]\,,
\label{sc-gb}
\end{equation}
where $r_H=\mu^{1/(D-3)}$ is the black hole horizon radius, and $r_{0}$, as
mentioned above, the value of the radial coordinate where the effective
potential attains its maximum value. For scalar, Dirac,
and electromagnetic perturbations, this is given by

\begin{equation}   
r_{0} = \left(\frac{D-1}{2}\right)^{\frac{1}{D-3}} r_{H} + O(1/\ell)\,.
\label{r0-Sch} 
\end{equation}        
Note that, for electromagnetic perturbations, the above equation is an exact one,
valid for any multipole number $\ell$. An expression similar to Eq. (\ref{sc-gb})
can be found also for Dirac perturbations, and has the form

\begin{equation} 
\omega = \sqrt{\frac{r_{0}^{D-3} - r_{H}^{D-3}}{r_{0}^{D-1}}}\,\left[k 
- i \sqrt{D-3} \left(n + \frac{1}{2} \right) \right]\,.
\label{dirac}
\end{equation}   
One may easily note that the above formulae, valid for large multipole
numbers, work well even for relatively small $\ell$, being a good approximation
already at $\ell=4$.   

\begin{widetext}

\begin{table}[t]
\caption{WKB values of the quasi-normal frequencies for scalar field
perturbations in the 3rd and 6th order beyond the eikonal approximation
with $\ell \geq  n$, for a $D$-dimensional Schwarzschild black hole
projected on the brane. }  
\label{Schwarz-sc}
\begin{ruledtabular}
\begin{tabular}{cccccc}
$D=5$ & WKB3  & WKB6 & $D=6$ & WKB3 & WKB6 \\ 
\hline \\
$\ell=0$; n=0;  & 0.265391 - 0.497380 i  & 0.256429 - 0.391293 i  & $\ell=0$; n=0; &
0.249904 - 0.793263 i   &  0.146960 - 0.600541 i \\
$\ell=1$; n=0;  & 0.722899 - 0.368853 i  & 0.748461 - 0.371463 i  & $\ell=1$; n=0; 
& 0.705383 - 0.527453 i   & 0.834846 - 0.512377 i \\
$\ell=1$; n=1;  & 0.535200 - 1.234129 i  & 0.568185 - 1.214473 i & $\ell=1$; n=1; 
& 0.355372 - 1.918880 i   & 0.392642 - 1.711230 i  \\ 
$\ell=2$; n=0;  & 1.243914 - 0.358055 i  & 1.249839 - 0.358066 i & $\ell=2$; n=0;  
& 1.376420 - 0.499199  i  & 1.393460 - 0.507541  i  \\ 
$\ell=2$; n=1;  & 1.099791 - 1.124086 i  & 1.111651 - 1.121364 i & $\ell=2$; n=1; 
& 1.039578 - 1.636385  i  & 1.088113 - 1.656560  i  \\ 
$\ell=2$; n=2;  & 0.887537 - 1.946290 i  & 0.887876 - 2.018084 i & $\ell=2$; n=2;  
& 0.564026 - 2.898250 i   & 0.568158 - 3.208780 i \\
$\ell=3$; n=0;  & 1.747954 - 0.355789 i  & 1.750048 - 0.355587 i & $\ell=3$; n=0;  
& 1.971510 - 0.497366  i  & 1.978550 - 0.496292 i \\ 
$\ell=3$; n=1;  & 1.642277 - 1.091913 i  & 1.646819 - 1.090268 i & $\ell=3$; n=1; 
& 1.733260 - 1.551881  i  & 1.752046 - 1.543178 i  \\ 
$\ell=3$; n=2;  & 1.468478 - 1.870707 i  & 1.458460 - 1.895964 i & $\ell=3$; n=2;  
& 1.347778 - 2.703307 i   & 1.317857 - 2.782190  i  \\ 
$\ell=3$; n=3;  & 1.246835 - 2.677609 i  & 1.219431 - 2.814159 i & $\ell=3$; n=3;  
& 0.851976 - 3.917659  i  & 0.742520 - 4.384462  i  \\
\end{tabular} 
\end{ruledtabular}
\end{table}

\begin{table}
\caption{WKB values of the quasi-normal frequencies for Dirac field
perturbations in the 3rd and 6th order beyond the eikonal approximation
with $k \geq  n$, for a $D$-dimensional Schwarzschild black hole
projected on the brane. }  
\label{Schwarz-fer}
\begin{ruledtabular}
\begin{tabular}{cccccc}
  $D=5$ &  WKB3  & WKB6 & $D=6$ & WKB3  & WKB6 \\ 
\hline \\
k=1; n=0;  & 0.373020 - 0.400822 i   &    0.427813 - 0.324042 i & k=1; n=0; 
& 0.296822 - 0.633767 i &  0.389272 - 0.404875  \\
k=1; n=1;  & 0.236229 - 1.299335 i   &    0.199685 - 1.251460 i & k=1; n=1; 
& 0.086481 - 1.975351 i &  -- \\ 
k=2; n=0;  & 0.945819 - 0.355450 i   &    0.973434 - 0.351309 i & k=2; n=0; 
& 1.000893 - 0.499552 i &  1.122648 - 0.454934 i\\ 
k=2; n=1;  & 0.771922 - 1.163379 i     &  0.806766 - 1.115347 i & k=2; n=1; 
& 0.664689 - 1.765862 i &  0.712367 - 1.523422 i \\ 
k=2; n=2;  & 0.555975 - 2.029365 i     &  0.580093 - 2.075602 i & k=2; n=2; 
& 0.280211 - 3.100380 i  & 0.329359 - 3.257230 i  \\
k=3; n=0;  & 1.471906 - 0.353568 i     &  1.479067 - 0.354610 i & k=3; n=0; 
& 1.641846 - 0.490566 i  & 1.668661 - 0.502019 i  \\ 
k=3; n=1;  & 1.341192 - 1.101171 i     &  1.359844 - 1.095934 i & k=3; n=1; 
& 1.334248 - 1.595761 i &  1.417715 - 1.561658 i \\ 
k=3; n=2;  & 1.143877 - 1.905989 i     &  1.152074 - 1.932798 i & k=3; n=2; 
& 0.913705 - 2.836118 i &  0.932631 - 2.876656 i  \\ 
k=3; n=3;  & 0.898975 - 2.738309 i     &  0.906727 - 2.920396 i & k=3; n=3; 
 & 0.392077 - 4.126275 i  & 0.331583 - 4.842823 i  \\
\end{tabular} 
\end{ruledtabular}
\end{table}

\begin{table}
\caption{WKB values of the quasi-normal frequencies for electromagnetic field
perturbations in the 3rd and 6th order beyond the eikonal approximation
with $\ell \geq  n$, for a $D$-dimensional Schwarzschild black hole
projected on the brane. } 
\label{Schwarz-gb}
\begin{ruledtabular}
\begin{tabular}{cccccc}
 $D=5$ & WKB3  & WKB6 & $D=6$ & WKB3 & WKB6 \\ 
\hline \\
$\ell=1 $; n=0;  & 0.545243 - 0.331346 i   & 0.576508 - 0.313923 i & $\ell=1 $; n=0;
&  0.482332 - 0.459270 i &    0.624699 - 0.303498 i \\
$\ell=1 $; n=1;  & 0.317482 - 1.135031 i   & 0.339969 - 1.081171 i & $\ell=1 $; n=1;
&  0.035275 - 1.686996 i &    --\\ 
$\ell=2 $; n=0;  & 1.142703 - 0.343905 i   & 1.148267 - 0.342292 i & $\ell=2 $; n=0;
&  1.244327 - 0.472984 i &    1.261845 - 0.461549 i\\ 
$\ell=2 $; n=1;  & 0.992174 - 1.082547 i   & 1.001222 - 1.074150 i  & $\ell=2 $; n=1;
&  0.917401 - 1.543831 i &    0.937130 - 1.486197 i \\ 
$\ell=2 $; n=2;  & 0.767162 - 1.877775 i   & 0.755782 - 1.948901 i & $\ell=2 $; n=2; 
&  0.437738 - 2.732445 i &    0.351770 - 2.942018 i \\
$\ell=3 $; n=0;  & 1.675991 - 0.348490 i   & 1.677917 - 0.348103 i & $\ell=3 $; n=0; 
&  1.874058 - 0.482782 i &    1.879429 - 0.480246 i\\ 
$\ell=3 $; n=1;  & 1.567818 - 1.070305 i   & 1.571663 - 1.067999 i & $\ell=3 $; n=1; 
&  1.637735 - 1.506514 i   &  1.649296 - 1.490905 i \\ 
$\ell=3 $; n=2;  & 1.389311 - 1.835125 i   & 1.376476 - 1.860555 i & $\ell=3 $; n=2; 
&  1.252908 - 2.624888 i   &  1.201152 - 2.690547 i \\ 
$\ell=3 $; n=3;  & 1.160459 - 2.628529 i   & 1.126499 - 2.770443 i & $\ell=3 $; n=3; 
&  0.755313 - 3.806577 i   &  0.604352 - 4.260917 i \\
\end{tabular} 
\end{ruledtabular}
\end{table}


\end{widetext}

By looking at our data, displayed in Tables I-III, one may draw important conclusions
for the dependence of the QN frequencies on the dimensionality of spacetime $D$,
spin $s$ of the particle and multipole moment $\ell$. Although the various SM 
fields are restricted to live on a 4-dimensional brane, their QN behaviour is
significantly different from the one in a purely 4-dimensional Schwarzschild
background, and resembles more the one of higher-dimensional fields living in
the bulk. As in the case of gravitons propagating in the higher-dimensional
spacetime \cite{RNd}, the imaginary part of the QN frequencies for all types
of brane-localised fields increase, as $D$ also increases. As a result, the
greater the number of hidden extra dimensions is, the greater the damping rate,
and thus the shorter-lived the quasi-normal ringing phase on the brane. The
behaviour of the real part of the QN frequencies is however less monotonic,
and seems to crucially depend on the values of the multipole moment $\ell$ and
overtone number $n$.
For $\ell, k > 2$, the fundamental ($n=0$) and first ($n=1$) overtones
are characterized by an increasing real oscillation frequency, as the value
of $D$ increases, in agreement with similar results derived for gravitational
QNMs in the bulk \cite{RNd}. We expect this behaviour to hold for
arbitrarily large multipole number where the eikonal approximation becomes
valid: Fig. 1 depicts the dependence of $\omega_{\rm Re}$ and $\omega_{\rm Im}$
as a function of $D$ in the eikonal regime, with the increase in both parts
being obvious. For $\ell, k \leq 2$, or $n \simeq \ell, k$ though, our data
seem to indicate that the real part of the QN frequency tends to decrease
with the total number of dimensions, with the only exceptions to this rule
being the scalar and electromagnetic fundamental mode ($n=0$) for $\ell=1$.
As nevertheless noted previously, the accuracy of our calculation decreases
exactly in this regime, therefore this result should be taken with caution
and confirmed by numerical analysis.

Turning to the dependence of the QNM frequencies on the spin of the particle,
let us consider the fundamental modes of all three types of SM fields that correspond
to the lowest possible value of the multipole number in each case. Within the 6th
order WKB approximation, we obtained:  $\omega_{n=0}= 0.25643 - 0.39129 i$ for scalar
($\ell=0$), $\omega_{n=0} = 0.42781 - 0.32404 i$ for Dirac ($k=1$), and
$\omega_{n=0} = 0.57651 -0.31392 i$ for gauge fields ($\ell=1$). Therefore,
among the lowest fundamental modes, the greater the spin of the field is, the
greater the $\omega_{\rm Re}$, and the smaller the $\omega_{\rm Im}$. 
As a result, fields with higher spin will decay slower and will dominate
during the ``final ringdown'' stage. To this respect, the brane-localised
fields behave in a similar way to purely 4-dimensional ones \cite{quasi-4D}. In
contrast, in studying gravitational perturbations in the bulk, it was found
that an increasing spin-weight led again to the enhancement of the real
frequency but to the increase of the damping rate as well \cite{RNd}.
Among the lowest three fundamental modes mentioned above, the field with
the higher spin ($s=1$) has also the largest ratio 
$|\omega_{\rm Re} /\omega_{\rm Im}|$, known as the quality factor of the
oscillator, a result that renders this mode the best oscillator of all three
in the vicinity of the projected-on-the-brane black hole.
However, if we look at the fundamental modes corresponding to higher values of the
multipole number, one may see that, for fixed $\ell$, it is instead the
scalar field that is the better oscillator of all three species. 

Finally, for fixed dimensionality of spacetime and spin of the particle, the
various QN frequencies still depend on the multipole number $\ell$. A simple
inspection of our data shows that, for brane-localised fields, the real
oscillatory part increases with $\ell$, for all species of fields and values
of $D$. The imaginary part of the frequency is decreasing for scalar fields
and increasing for fermions and gauge fields, however in all cases it soon
approaches a constant value. This is in agreement with Eqs. (\ref{sc-gb}) and
(\ref{dirac}), from where we may see that, in the high frequency approximation
($\ell \rightarrow \infty$), the real part of each mode is proportional to
the multipole number $\ell$, while the damping rate approaches a constant value
that depends solely on the fundamental parameters of the gravitational system
such as $r_H$, $r_0$ and $D$.


\section{Charged non-rotating black hole:  WKB values of Brane QNMs}

In this case, the metric function $h(r)$, that describes the projected-on-the-brane
Reissner-Nordstr\"om black-hole line-element, is given by Eq. (\ref{h-fun-RN}). 
The corresponding effective potentials $V_s$ for the various SM fields 
propagating on the brane are again determined by the formulae in Eqs. 
(\ref{spin-1})-(\ref{spin-0}), and, as before, have the form of positive-definite
potential barriers. As a result, the WKB method can be consistently applied also
in this case. 

The quasi-normal modes of a 4-dimensional Reissner-Nordstr\"om black hole were
investigated in \cite{RN1} within the WKB approximation, and in \cite{RN2}
with the help of the Frobenius method. Quasi-normal modes of more general, or
alternative, solutions for charged black holes were also considered in the literature: 
($D>4$)-dimensional Reissner-Nordstr\"om black holes \cite{RNd}, dilaton
black holes \cite{dilaton}, and Born-Infeld black holes \cite{Fernando}. 
In all the aforementioned works, one common feature was found to emerge
concerning the QNMs of charged black holes: the damping rate was monotonically
increasing as a function of the charge $Q$, until a critical value
$Q \approx 0.7-0.8 Q_{ext}$ was reached; after this point, the behaviour of
the damping rate changed into sharp decreasing. Thereby, $\omega_{\rm Im}$
as a function of $Q$ had a maximum somewhere near the value $Q \approx 0.7-0.8 Q_{ext}$,
where $Q_{ext}=\mu/2$ the value of the charge that corresponds to an extreme
charged black hole with a degenerate horizon.

In Figs. 2 \& 3 and Table IV, we display an indicative sample of our results,
derived by using the 6th order beyond the eikonal approximation WKB method;
these correspond to the fundamental modes ($n=0$) for scalars, fermions and
gauge fields, for the multipole number values $\ell=2$ and $\ell=1$, respectively.
For simplicity, the mass parameter and therefore the extremal value of the charge
have been fixed to the values $\mu=1$ and $Q_{ext}=0.5$, respectively. For a given
value of the charge $Q$, the dependence of the QNMs on $D$ and spin $s$ remains
the same as in the neutral case, therefore we will not comment further on this
dependence here. We will instead focus our attention to the dependence of the QN
frequency solely on the charge of the black hole.

According to our results, the quasi-normal modes of particles living in a charged
black-hole background projected on the brane have one, very distinctive feature
compared to the previously studied Reissner-Nordstr\"om cases\,: the characteristic
maximum of the damping rate as a function of the charge is absent. Indeed,
as one may see in Fig. 2, the damping rate of the $\ell=2$ fundamental modes of
all species of fields is a monotonically decreasing function of the charge $Q$. 
The same behaviour was found for the fundamental modes of all species corresponding
to higher multipole numbers, $\ell > 2$. The absence of the aforementioned maximum
is not a feature unseen before\,: it has been observed recently for $D$-dimensional
black holes in Gauss-Bonnet theory \cite{GB1}. The real oscillatory part of the
QN frequency on the other hand was found to be monotonically increasing, as a
function of $Q$, for all species of particles, and its exact dependence for the
case $\ell=2$ is depicted in Fig. 3.


\begin{figure}[t]
\resizebox{1\linewidth}{!}{\includegraphics*{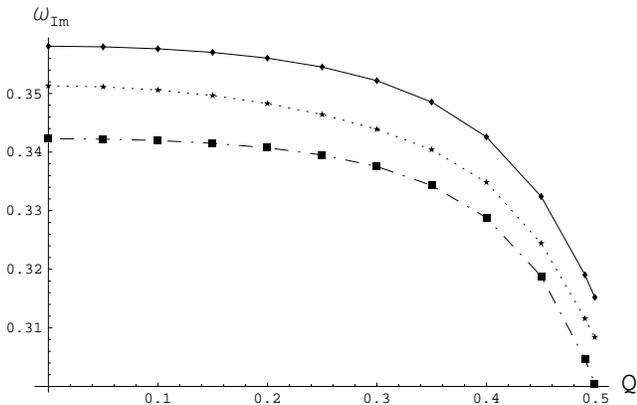}}
\caption{Dependence of imaginary part of  quasinormal modes on charge $Q$
for $\ell =2$; $r_{H}=1$, $D$ = 5. Scalar (diamond), Dirac (star),
and vector (box) perturbations.}   
\label{fig_2}
\end{figure}


As in the previous subsection, the fundamental modes of some species of
fields for the multipole number $\ell=1$ offer an exception to the
aforementioned rule. In Table IV, we display these modes for
all species of fields of the Standard Model. While the $\ell=1$ fundamental
mode of scalar fields follows the behaviour displayed in Figs. 2 and 3 both
for the imaginary and real part of the QN frequency, the one of fermions and
gauge bosons deviate by exhibiting a non-monotonic behaviour. In the case of
fermions, the imaginary part of the QN frequency does initially decrease
with the charge $Q$, as the rule dictates, however for higher values of $Q$,
it goes through a maximum
value before decreasing again; similarly, its real part starts increasing
as expected but again, after a critical point is reached, it starts decreasing.
In the case of gauge bosons, the real part of the $\ell=1$ fundamental mode 
does indeed follow the pattern shown in Fig. 3; however, its imaginary
part seems to follow instead the traditional Reissner-Nordstr\"om frequency 
behaviour by initially increasing with $Q$ and then decreasing. The behaviour
described above holds both for $D=5$ and $D=6$, with the increase in
the dimensionality of spacetime enhancing the height of the various
maxima points.


\begin{figure}[t]
\resizebox{1\linewidth}{!}{\includegraphics*{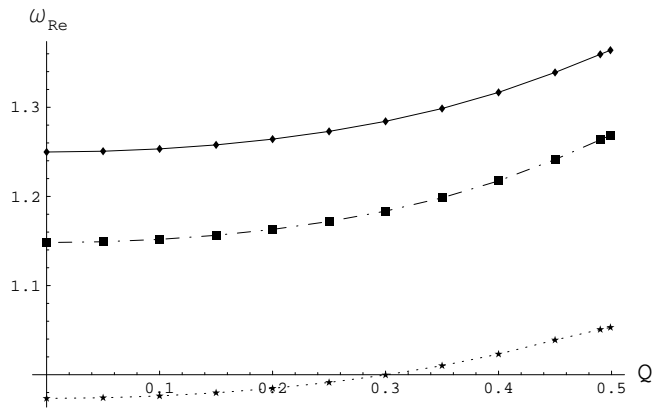}}
\caption{Dependence of real part of  quasinormal modes on charge $Q$
for $\ell =2$; $r_{H}=1$, $D$ = 5. Scalar (diamond), vector (box), and
Dirac (star) perturbations.}   
\label{fig_3}
\end{figure}


\begin{table}[b]
\caption{WKB values in 6th order beyond the eikonal
approximation for quasi-normal frequencies of brane-localised fields,
for $D=5$ and $D=6$ Reissner-Nordstr\"om black holes; $\ell=1$, $n=0$.}  
\label{RN-d5}
\begin{ruledtabular}
\begin{tabular}{cccc}
 $s$ & $Q$ & $D=5$  & $D=6$ \\ 
\hline \\
 0 & 0.05  & 0.749152 - 0.371256 i  &  0.835852 - 0.511642 i \\
 0 & 0.15  & 0.754757 - 0.369605 i  & 0.843518 - 0.506642 i  \\
 0 & 0.25  & 0.766514 - 0.366119 i & 0.858076 - 0.499811 i  \\ 
 0 & 0.35  & 0.786714 - 0.358488 i  &  0.887353 - 0.484023 i \\ 
 0 & 0.45  & 0.818792 - 0.334299 i  & 0.934330 - 0.425248 i  \\ 
 0 & 0.499 & 0.830394 - 0.313416 i  &  0.935737 - 0.396232 i  \\
 1/2 & 0.05  & 0.428119 - 0.323953 i & 0.389386 - 0.403475 i   \\
 1/2 & 0.15  & 0.430879 - 0.322740 i & 0.392510 - 0.388419 i \\
 1/2 & 0.25  & 0.438019 - 0.318018 i & 0.409388 - 0.346311 i \\ 
 1/2 & 0.35  & 0.447564 - 0.314646 i & 0.390580 - 0.382414 i \\ 
 1/2 & 0.45  & 0.458170 - 0.320913 i & 0.419673 - 0.437986 i \\ 
 1/2 & 0.499 & 0.450377 - 0.290551 i & 0.355861 - 0.374075 i \\
 1 & 0.05 &    0.577071 - 0.314252 i & 0.625394 - 0.305570 i \\
 1 & 0.15  &   0.581787 - 0.316804 i & 0.632436 - 0.321188 i \\
 1 & 0.25  &   0.592555 - 0.321297 i & 0.655043 - 0.345745 i \\ 
 1 & 0.35  &   0.613317 - 0.324651 i & 0.712217 - 0.358226 i \\ 
 1 & 0.45  &   0.652107 - 0.311345 i & 0.811198 - 0.319470 i \\ 
 1 & 0.499 & 0.668226 - 0.289308 i   & 0.832820 - 0.291253 i \\
\end{tabular} 
\end{ruledtabular}
\end{table}

Let us note at this point that, throughout this analysis, we consider
minimally-coupled fields propagating in a given black-hole background
projected on the brane. As a result, we do not take into account the
interaction of the spin-1 propagating particle with the electromagnetic
field of the black-hole background. In a more realistic description,
the electromagnetic perturbations will induce gravitational ones and
vice verca, and the quasi-normal spectrum of vector perturbations may
be considerably different from the one described here.

As in the case of a neutral Schwarzschild black hole, an expression may
be derived that describes the QN frequency in the eikonal approximation,
i.e. in the limit of large multipole number. By making use of the first
order WKB approximation, one can obtain, in the eikonal regime
$\ell \rightarrow \infty$, the following formula  
\begin{eqnarray} 
&~& \hspace*{-1.5cm}\omega = \sqrt{\frac{Q^{2} - \mu\,r_0^{D-3} + r_{0}^{2
D-6}}{r_{0}^{2 D-4}}} \nonumber \\ &~& \hspace*{1cm}\times  \left[\ell + \frac{1}{2} 
- i C \left(n + \frac{1}{2} \right) \right]\,,
\end{eqnarray}  
for scalar and vector perturbations, and   
\begin{equation} 
\omega = \sqrt{\frac{Q^{2} - \mu\,r_{0}^{D-3} + r_{0}^{2 D-6}}{r_{0}^{2 D-4}}}\,\left[k 
- i C \left(n + \frac{1}{2} \right) \right]\,,
\end{equation}   
for Dirac perturbations. As before, $r_0$ denotes the value of the radial
coordinate that corresponds to the maximum of the effective potential, which
is now given by
\begin{equation}
r_{0} =\left[\frac{(D-1)\mu +  \sqrt{(D-1)^{2}
\mu^2 -16 (D-2) Q^{2}}}{4}\right]^{\frac{1}{D-3}}\,,       
\end{equation}
also in the limit $\ell \rightarrow \infty$.
In the above equations, $C$ is a rather cumbersome constant depending on the black
hole parameters $\mu$ and $Q$ and on the spacetime dimensionality $D$. When $Q = 0$,
the above formulae reduce to Eqs. (\ref{sc-gb})-(\ref{dirac}) for a neutral black hole.
One may clearly see that, even for a non-vanishing value of $Q$, the real part of 
the QN frequency increases proportionally to the multipole number, while the 
imaginary part, and thus the damping rate, is determined through $C$ by the fundamental
parameters of the problem.


\section{Schwarzschild-de Sitter black hole: WKB values of Brane QNMs}

We now turn to the case of a $D$-dimensional Schwarz\-schild-de Sitter (SdS) black hole
projected onto a 4-dimensional brane. The corresponding line-element is defined
in terms of the metric function $h(r)$ given in Eq. (\ref{h-fun-dS}) with a positive
cosmological constant, $\Lambda>0$. This gravitational background is characterized
by two horizons, the black hole event horizon $r_H$ and the cosmological horizon
$r_c$. The effective potentials (\ref{spin-1})-(\ref{spin-0}), again, take the
form of a potential barrier that vanishes at both horizons. We can, 
therefore, once again apply consistently the WKB method.

In Table V, we display our results for the QNM frequencies for scalar, Dirac
and electromagnetic perturbations propagating in the vicinity of a
projected-on-the-brane SdS black hole. The data correspond to the fundamental
modes for the value $\ell=1$ of the multipole number, and for the dimensionalities
$D=5$ and $D=6$. The value of the cosmological constant is parametrised in
terms of its extreme value $\Lambda_{ext}$, that corresponds to the
degenerate case where the black hole and cosmological horizon coincide.
This maximal value of $\Lambda$ is defined through the relation:  
\begin{equation}
\mu^2 = \frac{4 (D-3)^{D-3}}{(D-1)^{D-1}}\left(\frac{(D-1)(D-2)}
{2 \kappa^2_D \Lambda_{ext}}\right)^{D-3}\,.
\end{equation}
The behaviour of the QN frequencies of the various species of brane-localised
fields does not yield any surprises in regard with their dependence on the
cosmological constant. Similarly to the behaviour exhibited by fields in the
vicinity of a purely 4-dimensional SdS spacetime \cite{quasi-4D} or by
gravitational fields living in a $D$-dimensional SdS spacetime \cite{RNd},
both the real and imaginary parts of the QN frequency are decreasing, as
the value of the cosmological constant increases. This decrease is observed,
and has the same magnitude, for all species of fields: as $\Lambda$ ranges
from zero to its maximum value $\Lambda_{ext}$, all fields undergo a suppression
of their QN frequency by approximately 90\%. For any given value of $\Lambda$,
an increase in the dimensionality of spacetime results, as in the case of QNMs
of bulk gravitons, to an increase of the value of both $\omega_{\rm Re}$ and
$\omega_{\rm Im}$, nevertheless the relative suppression of $\omega$ with
$\Lambda$ remains the same.

\begin{table}[t]
\caption{WKB values in 6th order beyond the eikonal
approximation for quasi-normal frequencies of brane-localised fields,
for $D=5$ and $D=6$ SdS black holes; $\ell=1$, $n=0$.}  
\label{SDS}
\begin{ruledtabular}
\begin{tabular}{cccc}
  $s$ & $\Lambda /\Lambda_{ext}$ &  $D=5$  & $D=6$ \\ 
\hline \\
0 & 0.1  & 0.702057 - 0.358774 i & 0.775357 - 0.504744 i \\
0 & 0.4  & 0.545649 - 0.305565 i  & 0.588540 - 0.445438 i  \\
0 & 0.6  & 0.424729 - 0.251384  i & 0.450599 - 0.369360 i  \\ 
0 & 0.8  & 0.282756 - 0.172529 i &  0.291411 - 0.252903 i \\ 
0 & 0.99 & 0.061394 - 0.035506 i  &  0.064318 - 0.0495684 i  \\ 
1/2 & 0.1  & 0.413557 - 0.305111 i & 0.391439 - 0.375808 i   \\
1/2 & 0.4  & 0.357933 - 0.246785 i & 0.376180 - 0.298955 i \\
1/2 & 0.6  & 0.303241 - 0.204146 i & 0.335594 - 0.252227 i \\ 
1/2 & 0.8  & 0.201820 - 0.162479 i & 0.205018 - 0.234865  i \\ 
1/2 & 0.99  & 0.048920 - 0.035784 i & 0.054487 - 0.050475  i \\ 
1 & 0.1 &    0.550719 - 0.300547 i & 0.591162 - 0.306145 i \\
1 & 0.4  &   0.459051 - 0.253351 i & 0.483397 - 0.295120 i \\
1 & 0.6  &   0.379611 - 0.212105  i & 0.397151 - 0.265749 i \\ 
1 & 0.8  &   0.271415 - 0.154087 i & 0.283191 - 0.205254 i \\ 
1 & 0.99  &  0.061232 - 0.0353454 i & 0.063908 - 0.049419 i \\ 
\end{tabular} 
\end{ruledtabular}
\end{table}

As it is well known, in the regime of near extremal values of $\Lambda$, the
effective potential approaches the Poschl-Teller potential \cite{CardosoSdS}
for which there is an exact analytical solution. Therefore, the exact formula
for QNMs in the extremal limit is:
\begin{equation}
\frac{\omega}{k_{H}} = \sqrt{ \frac{\ell (\ell + 1) (r_{c} - r_{H})}{2
k_{H} r_{H}^{2}} -\frac{1}{4} } - i \left(n + \frac{1}{2} \right), 
\end{equation}
for scalar and electromagnetic perturbations, and 
\begin{equation}
\frac{\omega}{k_{H}} = \sqrt{ \frac{k^{2} (r_{c} - r_{H})}{2
k_{H} r_{H}^{2}} -\frac{1}{4} } - i \left(n + \frac{1}{2} \right), 
\end{equation}
for Dirac perturbations. Here, the radii of the event and cosmological horizon are
approximated by the formulae\,:
\begin{equation}
r_{H} \approx  \sqrt{\frac{(D-1)(D-2)}{2 \kappa^2_D \Lambda}} \left(\sqrt{\frac{D-3}{D-1}}  -
\frac{\delta}{D-1} \right),
\end{equation}

\begin{equation}
r_{c} \approx  \sqrt{\frac{(D-1)(D-2)}{2 \kappa^2_D \Lambda}} \left(\sqrt{\frac{D-3}{D-1}}  +
\frac{\delta}{D-1} \right),
\end{equation}    
respectively, where 
\begin{equation}
\delta = \sqrt{ 1- \frac{\mu^{2}(D-1)^{D-1}}{4 \left(\frac{(D-1)(D-2)}
{2 \kappa^2_D \Lambda}\right)^{D-3}(D-3)^{D-3}} }.
\end{equation}  
Finally, the surface gravity $k_H$ at the event horizon is given by the formula 
\begin{equation}
k_{H} = \frac{1}{2} \frac{d h(r)} {d r^{*}}|_{r = r_{H}}\,.
\end{equation}

As we can see from the above formulae, in the extremal limit, fields of
different spin decay with exactly the same rate. Indeed, from the entries
of Table V, one can see that, despite its initially different values,
$\omega_{\rm Im}$ reduces to almost the same value for all species of fields
when $\Lambda = 0.99 \Lambda_{ext}$.
Note that the same behaviour is observed in the QNM spectrum for ordinary
4-dimensional SdS black holes \cite{Zhidenko}, and is therefore not an
unexpected property of the QN spectrum of brane-localised fields.

In the regime of large multipole numbers $\ell$, the value of the
radial coordinate $r= r_{0}$, at which the effective potential has
a maximum, is again expressed by Eq. (\ref{r0-Sch}) for fields of any spin. 
Using the first order WKB formula, one can find 
\begin{eqnarray}
&~& \hspace*{-1.5cm}\omega = \sqrt{\frac{1}{r_{0}^{2}} -\frac{\mu}{r_{0}^{D-1}}
-\frac{2 \kappa^2_D \Lambda}{(D-1)(D-2)}} \times \nonumber \\
&~& \hspace*{2,5cm} \left[\ell +  \frac{1}{2} + i C (n + \frac{1}{2})\right],
\end{eqnarray} 
for scalar and electromagnetic perturbations, and 
\begin{equation}
\omega = \sqrt{\frac{1}{r_{0}^{2}} -\frac{\mu}{r_{0}^{D-1}}
-\frac{2 \kappa^2_D \Lambda}{(D-1)(D-2)}}\,\left[k + i C (n + \frac{1}{2}) \right],
\end{equation} 
for Dirac perturbations.
When $D=4$, the above formulae go over to the ones for SdS black holes 
presented in \cite{Zhidenko}.


\section{Schwarzschild-Anti de Sitter black hole: WKB values of Brane QNMs}

A great interest in quasi-normal modes for asymptotically Anti de Sitter (AdS) 
black holes was stipulated the last 5 years after the observation was made, first 
in \cite{Horowitz}, that quasi-normal modes of classical ($d+1$)-dimen\-sio\-nal
asymptotically AdS black holes should coincide with poles of the retarded Green
function in the dual conformal field theory (CFT) at finite temperature in $d$
dimensions in the limit of strong coupling. In this context, the black hole
temperature corresponds to the temperature in the dual thermal field theory.
Therefore, by making use of this correspondence, one can investigate different
thermal phenomena in the limit of strong coupling in CFT, such as dispersion
relations and the hydrodynamic limit of CFT \cite{Starinets}. 

Assuming that the background in the bulk coincides with the one of a Schwarzschild-Anti
de Sitter (SAdS) black hole, the projected-on-the-brane line-element will be described
once again by Eq. (\ref{metric-br}), where $h(r)$ is given in Eq. (\ref{h-fun-dS})
with the cosmo\-logical constant now taking negative values, $\Lambda < 0$.
For later use, we define the ``effec\-tive'' anti-de Sitter radius $R$ as
\begin{equation}  
R=\sqrt{\frac{(D-1) (D-2)}{2 \kappa^2_D | \Lambda |}}\,.
\label{AdS-radius}
\end{equation}

In the case of an SAdS background, the effective potentials (\ref{spin-1})-(\ref{spin-0})
vanish at the black hole horizon but are divergent at infinity, and therefore 
Dirichlet boundary conditions, demanding the vanishing of the fields themselves
at infinity, are physically motivated. For perturbations of scalar fields,
the application of the Dirichlet boundary conditions poses no problem, as these
are also required by the AdS/CFT correspondence. However, for fields of higher
spin, one needs to impose the boundary conditions not on the wave function
$\Psi$ of the field, but on some alternative function that has the correct
interpretation in the context of the dual CFT \cite{Starinets}. The form of
the correct boundary conditions for higher-spin fields in an
asymptotically AdS spacetime is a controversial, and still open, question.
Since our interest in the AdS case is motivated mainly by the AdS/CFT
interpretation of the quasi-normal modes, here we limit our discussion to
the case of scalar fields for which the correct boundary conditions are known.

In order to find the quasi-normal modes for scalar field perturbations in the
projected-on-the-brane SAdS background, we shall apply the Horowitz-Hubeny
method \cite{Horowitz}. As this method is described in many papers, here we
will outline only its main points.  Near the event horizon, one can expand
the wave function $\Psi(r)$ of the field in the following way:
\begin{equation}
\Psi(x) = \sum_{m=0}^{\infty} a_{m} (x-x_{H})^{m}, \quad x_{H}=1/r_{H}\,. 
\label{sum}
\end{equation}


\begin{widetext}

\begin{table}[t]
\caption{Fundamental quasi-normal frequencies ($\ell=0$, $n=0$) for scalar field
perturbations for a Schwarzschild-Anti de Sitter black hole; $D=5$, 6, and 7. }  
\label{AdS}
\begin{ruledtabular}
\begin{tabular}{cccc}
$r_{H}$ &  $D=5$  &  $D=6$  &     $D=7$     \\ 
\hline \\
100 $R$  &  262.02934 - 230.20913 i   & 299.44402 - 200.49039 i   & 320.32124 -  179.05396 i  \\
50 $R$  &  131.03212 - 115.10340 i     & 149.74331 - 100.24219 i   & 160.18552 - 89.52188 i \\
10 $R$  &  26.33589 - 23.00814 i      & 30.08465 - 20.02929 i   & 32.17341 - 17.87798 i   \\ 
5 $R$  &   13.36830  - 11.48460 i      & 15.25271 - 9.98509 i   & 16.30117 - 8.90885 i \\ 
1 $R$  &   3.7449 - 2.1871 i      &  4.17056 - 1.84684 i  & 4.39542 - 1.62536 i    \\ 
0.8 $R$  &   3.4051 - 1.7056 i       & 3.76193 - 1.42408 i &  3.94602 - 1.24788 i    \\
0.6 $R$  &   3.1239 - 1.2168 i        &  3.41142 - 0.99847  & 3.55367 - 0.87043 i   \\ 
\end{tabular}
\end{ruledtabular}
\end{table}

\end{widetext}
\noindent
Here, $r_{H}$ is the largest of the zeros of the metric function $h(r)$, that
corresponds to the black hole horizon radius. On the other hand, at infinity,
the Dirichlet boundary conditions, that we will use here, dictate that
\begin{equation}
|\Psi(r=+\infty)|\equiv |\Psi(x=0)| =0\,. \label{Dirichlet}
\end{equation}
The method amounts to solving the above equation, whose roots will then give us
the corresponding quasi-normal frequencies $\omega$. To this end, we need to
truncate the sum (\ref{sum}) at some large $m=N$, and make sure that the roots
of Eq. (\ref{Dirichlet}) still converge for higher values of $m$. Note that
the larger the dimensionality $D$ is, the larger the truncation number
$N$ needs to be. In addition, for small black holes and high overtones we need
a very large $N$ in order to find the QN modes with a good accuracy. Thus, for
instance, for $r_{H} \approx 0.6$ and $D=6$, $N \approx 10^{4}$.

\smallskip
The properties of the quasi-normal spectrum for a\-symp\-totically AdS black
holes strongly depend on the size of the black hole relative to the AdS
radius. Thus, QN spectra of large ($r_{H} \gg R$), intermediate ($r_{H}
\approx R$) and small ($r_{H} \ll R$)  black holes are totally different.
QNMs of large AdS black holes are proportional to the black hole radius
and therefore proportional to the temperature of the black hole
\cite{Horowitz}. In the regime of intermediate black holes, this
proportionality is broken \cite{Horowitz}, and, for small AdS black
holes, QNMs reduce to the normal modes of the AdS spacetime when a black hole
radius goes to zero \cite{konoplya2002(1)}. At high overtones, the
quasi-normal spectrum becomes equidistant, with a spacing which is
independent of the multipole number and spin of the field being
perturbed \cite{Cardoso-Konoplya-Lemos}.  Below, we shall show that 
the above properties also hold for a higher-dimensional, asymptotically
AdS black hole projected on the brane.


\begin{table}[b]
\caption{Higher overtones for scalar field perturbations of large
($r_{H} = 100 R$) SAdS black hole; $D=5$, $6$; $\ell = 0$}  
\label{GBAdS_L01_d5}
\begin{ruledtabular}
\begin{tabular}{cccc}
 $n$ &  $D=5$  &  $D=6$  &       \\ 
\hline \\
1  &  465.01779 - 431.64399 i & 541.12945 - 374.61983 i    &             \\
2  &  665.99215 - 632.02215 i  & 780.16118 - 547.73831 i   &        \\
3  &  866.46020 - 832.18457 i &  1018.54515 - 720.65433 i     &       \\ 
4  &  1066.73187 - 1032.27206 i &  1256.67421 - 893.49563 i &         \\ 
5  &  1266.90874 - 1232.32573 i  &   1494.67836 - 1066.30165 i   &         \\ 
6  &  1467.03327 - 1432.36158 i  &  1732.61261 - 1239.08844 i  &        \\
7  & 1667.12614 - 1632.38699 i  & 1970.50406 - 1411.86365 i      &        \\ 
8  & 1867.19851 - 1832.40582 i   & 2208.36754 - 1584.63141 i    &      \\ 
9  & 2067.25695 - 2032.420267 i    &  2446.21180 - 1757.39410 i    &          \\ 
\end{tabular} 
\end{ruledtabular}
\end{table}

The fundamental quasi-normal frequencies for scalar field perturbations
($\ell=0, n=0$) in the vicinity of an SAdS black hole projected on the brane,
are shown in Table VI, for the cases $D=5,6$ and 7. The black hole horizon radius
is taken to cover the regime from a fraction of the AdS radius to a hundred
times larger than $R$. From the displayed data, one may easily see that
the QN frequency increases as the black-hole horizon increases, and that,
for $r_{H} \gg R$, $\omega$ is indeed proportional to $r_{H}$. The asymptotic
regime of high damping begins at relatively small overtone numbers, i.e. at
$n \approx 10$. In Table VII, we display the first 9 overtones for a
projected-on-the-brane SAdS black hole with $r_H=100 R$, from where we may
see that the quasi-normal spectrum very quickly becomes equidistant
with a spacing that depends only on the dimensionality of spacetime $D$
and the black-hole horizon $r_{H}$. From the numerical data of Table VII,
we may deduce the spacing for both the real and imaginary part of the QN
frequency, that come out to be
\begin{equation}
\omega_{n+1} - \omega_{n} \approx 2 (1 + i)\,r_{H}, \quad D = 5,
\end{equation}    
\begin{equation}
\omega_{n+1} - \omega_{n} \approx (2.4 + 1.7 i)\,r_{H}, \quad D = 6.
\end{equation}

Let us consider next the limit of very small black-hole radius. 
From Ref. \cite{konoplya_smallAdS} one can learn, that the QNMs
of an asymptotically AdS black hole approaches its pure AdS
values (normal modes). The normal modes for scalar fields in an AdS
spacetime can be found analytically. For the case of a field propagating
on a brane, that is embedded in a pure AdS spacetime, the metric
function takes the form
\begin{equation}
h(r) = 1 - \frac{2 \kappa^2_D \Lambda r^{2}}{(D-1) (D-2)}, \quad \Lambda < 0\,,
\end{equation} 
and describes an ``effective'' 4-dimensional Anti de Sitter spacetime with
a re-scaled cosmological constant: $\Lambda \rightarrow 6 \Lambda/(D-1)(D-2)$. 
Therefore, the normal modes for brane-localised scalar field perturbations
can follow immediately from the corresponding analysis of the purely
4-dimensional case \cite{Cardoso-Konoplya-Lemos}, and are given by:
\begin{equation}
\omega_{scalar} = \sqrt{\frac{2 \kappa^2_D |\Lambda|}{(D-1) (D-2)}}\,(2 n + \ell + 3).
\label{normal}
\end{equation} 
For instance, for the particular case of $n=0$ and $D=5$, the above formula
leads to the result: $\omega_{n=0} \approx 3$. Then, from the entries of
Table VI, one may clearly see that, as the black-hole radius decreases,
the QN frequency asymptotes to the same value. Thus, the QN
modes of scalar fields, in the vicinity of a very small asymptotically AdS
black hole projected on the brane, approach indeed the normal modes of the
pure AdS spacetime.


\begin{figure}
\resizebox{1\linewidth}{!}{\includegraphics*{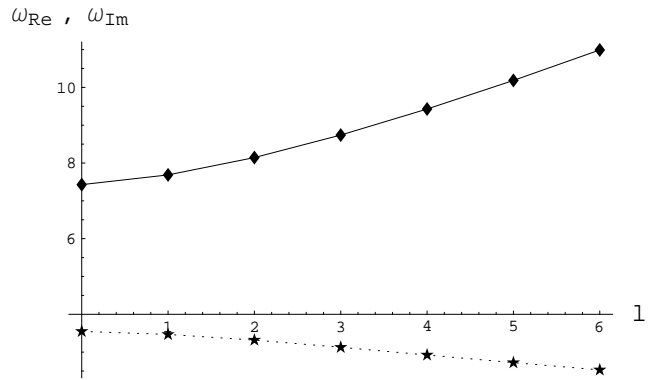}}
\caption{Dependence of the real (diamond) and imaginary (star) part
of QNMs on multipole number  $\ell$, for $n=0$, $r_{H}=1$ and $D$ = 6.
Scalar field perturbations of SAdS black hole.}   
\label{fig_4}
\end{figure}


In Fig. 4, we depict the dependence of the QNM frequency spectrum on
the multipole number $\ell$. In constrast to what happens in the
asymptotically flat case, where the imaginary part of the QN frequency
was approaching a constant value in the limit $\ell \rightarrow \infty$,
in the case of a SAdS black hole, the damping rate is decreasing when
$\ell$ is growing, therefore higher multipoles should decay more slowly.
Yet, as was mentioned in \cite{Horowitz}, this does not pose a problem
for AdS/CFT correspondence since a decomposition according to multipole
numbers can be done in the dual CFT as well.

Finally, another deviation -- from the behaviour noted in the previously 
studied cases -- is observed here, this time in respect to the dependence
of the QN spectrum on the dimensionality of spacetime. According to our
results, in the case of a projected-on-the-brane SAdS black hole, when
the number of hidden, transverse dimensions increases, the imaginary part
of the QN frequencies is decreasing, while the real part of $\omega$ is
increasing. We remind the reader that in the case of asymptotically flat
or de Sitter black holes, both  $\omega_{\rm Re}$, and $\omega_{\rm Im}$ 
grow when $D$ is increasing (see, for example, Fig. 1).

\section{Conclusions}

The theories postulating the existence of extra, compact spacelike dimensions
have opened the way for low-scale gravitational theories and the observation
of strong gravitational phenomena such as the creation of tiny, higher-dimensional
black holes in the controlled environment of a ground-based particle accelerator.
Standard Model fields, that are restricted to live on the 4-dimensional brane
and happen to propagate in the vicinity of such a black hole, will feel only the
projected-on-the brane black-hole background. In such a situation, the
observation of their QN frequencies may be highly more likely to take place
than the one for the diligently pursued, but up to now elusive, gravitons.
Such a detection will provide not only evidence for the existence of the black
holes themselves, but also information on the fundamental parameters of the
higher-dimensional theory.

In this work, we have investigated a variety of sphe\-ri\-cally-symmetric
$D$-dimensional black-hole backgrounds, that are then projected onto the
4-dimensional brane. The induced-on-the-brane backgrounds depend on a single
metric function $h(r)$, that carries a signature of the parameters
of the $D$-dimensional theory, such as the dimensionality $D$, charge $Q$ and
cosmological constant $\Lambda$. The same function characterizes the form of
the effective potentials felt by the scalars, fermions and gauge bosons 
propagating in the brane background. In 3 of the cases studied (Schwarzschild,
Reissner-Nordstr\"om and Schwarzschild-de Sitter), the effective potentials had
the form of positive-definite barriers, a result that allowed us to use the WKB
method to derive the QN spectrum in the 6th order beyond the eikonal approximation.
In the 4th case (Schwarzschild-Anti de Sitter), the effective potential diverged
at infinity, and the Horowitz-Hubeny method was used instead.

In the case of a Schwarzschild black-hole background, the QN spectrum of all
SM fields was computed for various values of the dimensionality of spacetime
$D$ and multipole number $\ell$.
It was found that, as the number of hidden, extra dimensions
increased, the imaginary part of the QN frequency for all species increased
as well, thus making the ring-down phase on the brane shorter. The real part
of the QN frequency was also found to be $D$-dependent and to predominanlty
increase with $D$, although particular modes with $\ell \simeq n$ may deviate
from this behaviour. In respect to the dependence on the spin, fields with
higher spin were found to damp with a slower rate and thus to survive longer.

If a charge $Q$ is present in the bulk background, the QN frequency spectrum 
is found to significantly deviate from the one of purely 4-dimensional ones,
and to resemble more the one in the vicinity of a $D$-dimensional Gauss-Bonnet
black hole: an increase in the charge of the black hole was found to lead to a 
monotonically decreasing behaviour for the imaginary part of the QN frequency 
-- and thus to a longer ring-down phase -- and to a monotonically increasing
behaviour for the real part, rendering all SM fields much better oscillators
than in the neutral case.

When a positive cosmological constant is turned on in the bulk, the resulting QN
spectrum of the SM fields on the brane does not yield any surprises. As $\Lambda$
increases, both the real and imaginary part of the QN frequency are suppressed,
and for $\Lambda=\Lambda_{ext}$, the suppression reaches the magnitude of 90\%.
The number of dimensions enhances again the individual values of $\omega_{\rm Re}$ 
and $\omega_{\rm Im}$, however the relative suppression as $\Lambda$ varies
comes out to be only mildly $D$-dependent. In the extreme limit, the feature
of fields with different spin decaying with almost the same rate is observed
also in the present case of brane-localised fields in an effective SdS background.

In the case of a negative cosmological constant being present in the bulk, the
spectrum of QN frequencies was shown to depend on the ratio of the black-hole
horizon to the AdS radius, similarly to the case of purely 4-dimensional or
$D$-dimensional SAdS backgrounds.  For large black holes, the QN spectrum comes
out to be proportional to $r_H$, and to become equidistant for the higher
overtones. In the opposite limit of a very small black hole, the QN frequencies
approach the normal modes of the projected-on-the-brane AdS spacetime.

Summarizing our results, we may say that the existence of extra dimensions
affects the QNM spectrum of fields living on the brane both in a direct and
an indirect way. In all the cases studied here, an increase in the number of
transverse-to-the-brane dimensions causes a significant increase in the imaginary
part of all fields, thus directly affecting the damping rate of the field
perturbations on the brane. Additional features, such as the distance in the
frequency spacing between successive quasi-normal modes in the SAdS spacetime,
or asymptotic values of QN frequencies in various limits and backgrounds, also
depend explicitly on the total number of dimensions $D$. In an indirect way, 
the localization of fields on a brane embedded in a higher-dimensional black-hole
background leads to a deviation from the behaviour observed either in the purely
4-dimensional or in a purely $D$-dimensional case: the monotonic behaviour of
the QN spectrum as a function of the charge $Q$, for the majority of the modes, 
is an indicative example of this.

In this work, we have restricted our analysis to the study of spherically-symmetric
$D$-dimensional black-hole backgrounds projected on the brane. The study of
axi\-ally-symmetric, rotating black-hole backgrounds has already been initiated and
we hope to report our results soon in a follow-up article. In addition, here we
were limited by the consideration of quasi-normal spectra for only massless fields,
yet, as was shown in \cite{massivescalar} for scalars, and in \cite{massivevector}
for vector fields, the mass term can change the lower modes of the spectrum
considerably, leaving unaffected the high damping limit of the spectrum -- the
study of the effect of the mass term on the QN spectrum of brane-localized fields
in various backgrounds is also among our future plans.
\vspace{3mm}


\begin{acknowledgments}

P.K. acknowledges financial support from the U.K. Particle Physics and Astronomy 
Research Council (Grant Number PPA/A/S/2002/00350).
The work of R. K. was supported by \emph{Funda\c{c}\~{a}o de Amparo
\`{a} Pesquisa do Estado de S\~{a}o Paulo (FAPESP)}, Brazil.

\end{acknowledgments}


%

%

\end{document}